%%%%%%%%%%%%%%%%%%%%%%%%%%%%%%%%%%%%%%%%%%%%%%%%%%%%%%%%%%%%%
% Dwarf Galaxies and the Origin of Intracluster Medium
%          Biman B. Nath and Masashi Chiba 
%                    1995:02:01
%%%%%%%%%%%%%%%%%%%%%%%%%%%%%%%%%%%%%%%%%%%%%%%%%%%%%%%%%%%%%

\newcount\fcount \fcount=0
\def\ref#1{\global\advance\fcount by 1
\global\xdef#1{\relax\the\fcount}}

\def\la{\lower.5ex\hbox{$\; \buildrel < \over \sim \;$}}
\def\ga{\lower.5ex\hbox{$\; \buildrel > \over \sim \;$}}
\def\r{\hangindent=1pc \noindent}
\def\etal{et~al.}

\font\large=cmr10 scaled\magstep1
\font\sc=cmr10 scaled\magstep0

\magnification=\magstep1

\raggedbottom
\footline={\ifnum\pageno=1 \hfil \else \hss\tenrm\folio\hss\fi}

\tolerance = 30000

\baselineskip=5.00mm plus 0.1mm minus 0.1mm
%%%%%%%%%%%%%%%%%%%%%%%%%%%%%%%%%%%%%%%%%%%%%%%%%%%%%%%%%%%%%

\centerline{\bf \large{DWARF GALAXIES AND THE ORIGIN OF INTRACLUSTER MEDIUM}}
\vskip 1.0cm
 \centerline{\large{B{\sc{IMAN}} B. N{\sc ATH$^1$} {\sc AND}
M{\sc ASASHI} C{\sc HIBA$^2$}} }
 \vskip 0.2cm
\centerline{$^{1}$Inter-University Center for Astronomy \& Astrophysics,
Post Bag 4, Pune 411007, India}
 \centerline{$^{2}$Astronomical Institute, Tohoku University,
Aoba-ku, Sendai 980-77, Japan}
\noindent

\vskip 1.0cm
\centerline{To appear in The Astrophysical Journal}
\vskip 0.5cm
\centerline{\it Received 10 February 1995; \quad accepted 
15 May, 1995}
\vskip 1.5cm

%%%%%%%%%%%%%%%%%%%%%%%%%%%%%%%%%%%%%%%%%%%%%%%%%%%%%%%%%%%%%%%%%%%%%%%
\centerline{ABSTRACT}
\medskip
Following the recent suggestion that it is dwarf galaxies in clusters
-- as opposed to large ellipticals -- that provide the intracluster gas, 
we estimate the metallicity of the intracluster medium
(ICM) in such a case. We derive analytical expressions for the fraction
of mass of dwarf galaxies that is ejected, and estimate the metallicity
of the resulting intracluster gas. We find that the metallicity resulting
from this hypothesis is adequate only for clusters with low-metallicity gas.
Since galactic winds from dwarf galaxies occur much earlier than
those from ellipticals, due to the smaller binding energy of the gas, we show
that the gas ejected by dwarf galaxies is enriched mostly by Type II 
supernovae, as opposed to Type I supernovae in the case of large galaxies.
We further point out that the gas in small scale structures, which
never cools and forms stars due to low temperatures and, consequently, large
cooling time scale, gets dispersed in the process of hierarchical
clustering, and is incorporated as the diffuse intracluster gas. We show 
that this process can provide enough hot gas to be compatible with X-ray
observations in rich clusters.
\bigskip

\noindent
{\it Subject headings:} Galaxies : abundances -- Galaxies : clustering -- 
Galaxies : evolution -- Galaxies : intergalactic medium
\vskip 1cm
%%%%%%%%%%%%%%%%%%%%%%%%%%%%%%%%%%%%%%%%%%%%%%%%%%%%%%%%%%%%%
%%% Section 1 %%%%%%%%%%%%%%%%%%%%%%%%%%%%%%%%%%%%%%%%%%%%%%%
%%%%%%%%%%%%%%%%%%%%%%%%%%%%%%%%%%%%%%%%%%%%%%%%%%%%%%%%%%%%%
\centerline{1. INTRODUCTION}
\medskip

The existence of hot gas in clusters of galaxies was established
from X-ray observations almost two decades ago, and so was its 
metallicity, especially from the emission lines of Fe, Si, S, Mg and O
(e.g., Sarazin 1988). 
The amount of the hot gas has been inferred to be a substantial fraction 
of the total gravitational mass of clusters (e.g., White \etal 1993).
For the origin of this gas, several hypothesis have been put forward, the
most pursued one being that of galactic winds from elliptical galaxies 
driven by supernovae (Arimoto \& Yoshii 1987; Matteucci \& Tornamb\`e
1987; David \etal 1990).
%from destruction of late-forming giant galaxies (Whitmore \etal 1993). 

Several authors have, however, found that the amount of gas 
expelled from the visible 
ellipticals and S0 galaxies is not adequate to account for the total gas 
mass (David \etal 1990; Matteucci and Vettolani 1988; Okazaki \etal 1993). A
large portion of the ICM gas,
even as large as $90\%$ of it, has therefore been thought to be
primordial. This interpretation, coupled with the fact that the ratio of the 
intracluster gas mass to the stellar mass in the galaxies is higher
in rich clusters, has led to the idea that galaxy formation is less efficient
in rich clusters (David \etal 1990; Matteucci and Vettolani 1988).

The metallicity of the ICM gas has been estimated by various authors 
from the hypothesis that the metals are produced and expelled in 
galactic winds from ellipticals. Chemical evolutionary models of 
ellipticals have been used to predict the metallicity (e.g., Arimoto \&
Yoshii 1987; Matteucci \& Tornamb\`e 1987). Okazaki \etal (1993)
found that these models are adequate to explain the observed 
total mass of iron in the Virgo cluster, but not to explain the total gas 
mass. Arnaud \etal (1992) defined a dimensionless ratio 
of iron mass to that of the stars in the galaxy, to analyze the data, and 
came to the conclusion that one needed a bimodal star formation
in cluster galaxies. Renzini \etal (1993) found this idea rather ad hoc,
and, introducing another ratio, that of the iron mass to blue light,
they found constraints from the data on the yields of Type I and II 
supernovae rate that
may explain the observations. It seems, therefore, that no single
model satisfactorily explains the iron mass (or the metallicity)
and the total gas mass of ICM, simultaneously.

Recently, Trentham (1994) has suggested that if dwarf galaxies, at the
lower end of the luminosity function, are numerous, i.e., if the index of
the Schechter luminosity function $\alpha \la -1.7$, as is seen in some 
clusters,
then they may supply enough gas. These galaxies, with shallow gravitational
potential wells, and with small binding energy of the gas, are vulnerable to
huge mass losses through winds driven by supernovae explosions
from the first generation of star formation (Larson 1974; Saito 1979).
Dekel \& Silk (1986) showed that galaxies with velocity dispersion less than
$\sim 100$ km s$^{-1}$ would lose much of their gas due to large galactic 
winds, and become low surface brightness dwarf galaxies. Trentham (1994)
argued that if a fraction $\gamma \sim 0.05 -0.33$ of the total mass of
the galaxy is expelled, and if the index of the Schechter luminosity function
$\alpha \sim (-1.9)-(-1.7)$, for masses $10^4 - 10^{11}$ M$_{\odot}$, then the
amount of gas mass expelled can account for the total gas mass in the ICM.

One of the tests of this interesting model lies in the resulting 
metallicity of the ICM gas. Since the galactic winds from dwarf galaxies
(of total mass $\la 10^{11}$ M$_{\odot}$) occur in a time scale $
t\la 10^9$ yr, Type I supernovae can not contribute substantially to the
metallicity. We suggest here that the hypothesis of dwarf
galaxies as sources of ICM gas leads necessarily to the dominance of Type II
supernovae as the origin of the metals in ICM. In this paper, we estimate the 
total metallicity of the ICM resulting from
galactic winds from dwarf galaxies using more physical values of $\gamma$,
rather than a universal number for all galaxies, and we derive the relations
between the metallicity and various parameters of star formation. 

The structure of the paper is as follows: In \S 2 we calculate the
galactic wind epoch for dwarf galaxies and the mass of iron that is
contributed by individual galaxies. In \S 3, the total mass of iron
in the intracluster gas is estimated using the mass function of 
galaxies. We discuss the nature of the metallicity of ICM in \S 4. 
In the paper, we write the present day Hubble constant as
$H_0=50 h_{50}$ km s$^{-1}$ Mpc$^{-1}$.

%%%%%%%%%%%%%%%%%%%%%%%%%%%%%%%%%%%%%%%%%%%%%%%%%%%%%%%%%%%%%
%%% Section 2 %%%%%%%%%%%%%%%%%%%%%%%%%%%%%%%%%%%%%%%%%%%%%%%
%%%%%%%%%%%%%%%%%%%%%%%%%%%%%%%%%%%%%%%%%%%%%%%%%%%%%%%%%%%%%
\bigskip\medskip
\centerline{2. GALACTIC WIND FROM DWARF GALAXIES}
\bigskip

Galactic winds are thought to be initiated when the thermal energy
of the gas in the parent galaxy exceeds its binding energy. Below, we
will consider galaxies with masses below $M_T \sim 10^{11}$ M$_{\odot}$,
where $M_T$ denotes the total mass. We will assume that
a fraction $(1-g)$ of the mass is in the form of dark matter, {\it i.e.},
the mass of gas ($M_g$) and that of stars ($M_s$) comprise a fraction $g$
of the total mass. A few million 
years after the onset of star formation,
the first generation of supernovae begins to explode (the life time of 
a $100$ M$_{\odot}$ star is $\sim 3.2 \times 10^6$ yr), and they put
energy into the interstellar medium. 

The energy in the supernovae blastwaves is turned into the thermal energy
of the gas, as the supernovae remnant (SNR) shells overlap and the thermal gas
inside the shells interacts with the ambient gas in the ISM.
(The energy in the dense shell is quickly radiated away and does not contribute
much in pumping energy into the ISM gas.) The radii of SNR expand as
(Cox 1972) 
$$\eqalignno{
R_{SNR} (t)&= R_0 (t/t_{rad})^{2/5}\>, \> \>\> {\rm for} \> t<t_{rad} &\cr
&=R_0 (t/t_{rad})^{2/7}\>, \> \>\> {\rm for} \> t>t_{rad}&(1)\cr}
$$
Here $R_0=0.039 \> n^{-0.4}$ kpc and $n$ is the particle density.
The total energy liberated in a typical supernovae is assumed
to be $10^{51}$ ergs. Half of the thermal energy of the SNR is radiated away
after a time $t_{rad}$, which, for primordial gas, is approximately equal to
$1.7 \times 10^5 n^{-0.49}$ yr for a primordial gas
(Babul \& Rees 1992). For a mildly enriched gas,
$t_{rad} \sim 0.9 \times 10^5 \lambda^{-5/17} n^{-9/17}$ (using a cooling rate
of $3 \times 10^{-18} \lambda T^{-1} n^2$ ergs cm$^{-3}$ s$^{-1}$, as in
Dekel \& Silk 1986), where $\lambda$ represents the metallicity of the gas
($\lambda=1$ for $Z=0.01$, and higher for larger value of $Z$). Since these
two time scales are within a factor of order unity, we will use the one for 
a primordial gas for simplicity.

If we define (as in Babul \& Rees 1992) $t_{snr}$, the time scale for the 
overlapping of SNRs, as the time when the remnant shells occupy $\sim 60 \%$
of the total volume, then $t_{snr}$ can be estimated in the following way.
We use for the radius of the galaxy, the expression derived by Saito (1979)
(or. e.g., Matteucci \& Tornamb\`e 1987), which is valid over a large range of 
masses, from
globular clusters to large ellipticals, {\it viz.},
$$
R= 1.2 \times 10^5 \Bigl ( { M_T  \over 10^{12} M_{\odot}} \Bigr )^{0.55}
\quad {\rm pc} \>,\eqno(2)
$$
where $M_T$ is the total mass of the elliptical galaxy.
(This is close to the estimate of the
radius of the starburst region of dwarf galaxies in Babul \& Rees 1992.)
We assume a constant star formation rate for analytic simplicity, and write 
the star formation rate as $\dot M _{\ast} \approx (\tau gM_T/t_{ff})$, 
where $t_{ff}=1.9 \times 10^7 g^{1/2} n^{-1/2}$ yr is the free fall time
of the gas, and $\tau (<1)$ is the efficiency of star formation.
We will see later that the final results of the metallicity of ICM do not
depend on $\tau$. The rate of supernovae
is given by the $\nu= 10^{-35} \nu _{50}$ per gram, where $\nu _{50}$ refers
to one supernova per 50 M$_{\odot}$ (see also, Dekel and Silk 1986). 
Therefore, the number of
supernovae in time $t$ is given simply by $N_{SN} (t)= \nu \dot M _{\ast} t$.

The rate of supernovae, as denoted by $\nu$, of course depends on the
slope of the initial mass function (IMF). For an IMF with slope $x$
[$\phi (m) \propto m^{-x} dm$], the factor $\nu _{50}$ can be written as,
$$
\nu _{50} = 50 ({ x-1 \over x}) \Bigl ({ 1 \over 
M_{l, sn}^x} -{ 1 \over M_u^{x}} \Bigr) \Bigl ( {1 \over M_l ^{x-1}} -
{ 1 \over M_u^{x-1}}
\Bigr )^{-1}\>, \eqno(3)
$$
where $M_l$ and $M_u$ are the lower and upper limits of masses of stars in
the main sequence, and $M_{l,sn}$ is the lower limit of the mass for the
progenitor star for a Type II supernova (all in the units of M$_{\odot}$).
For $M_l=0.1, M_u=100$, and $M_{l,sn}=8$, we get $\nu _{50} \sim 0.37$
(for $x=1.35$, Salpeter IMF), and $\nu_{50} \sim 0.90$ (for $x=0.95$).
The term $\nu _{50}$, therefore, succinctly expresses four important
parameters ($x, M_l, M_u, M_{l,sn}$) and it will be useful in comparing
our results later with the numerical results of previous workers.

The timescale of the SNR collision, $t_{snr}$, can then be estimated
as (in the units of $10^9$ yr),
$$
t_{snr} \approx 6 \times 10^{-3} n^{0.13} \Bigl ({\nu _{50} \over 0.35}
\Bigr )^{-0.54} \Bigl ({g \over 0.1} \Bigr )^{-0.27} \Bigl ( {M_T \over 
10^{12} \> M_{\odot}} \Bigr )^{0.35} \quad {\rm Gyr} \>. \eqno(4)
$$
Larson (1974) showed that the fraction of the energy of the blast ($=E_{SN}$) 
that is finally imparted to the gas is $\eta \sim 0.1$ (see also Babul 
\& Rees 1992), when the remnant shells overlap. We will see below that 
$t_{snr}$ in eqn ($4$) is much shorter than the time scale over which the 
galactic winds are excited.
Therefore, the assumption that $\eta \sim 0.1$ is valid,
since the whole galaxy would be pervaded by SNRs by the epoch of 
galactic winds.

As the continuing SNR collisions heat up the gas, the thermal energy content
of the gas ($E_{th}$) increases as,
$$
\eqalignno{
E_{th} (t) &=0 \>, \qquad \qquad \qquad \> t < 3.2 \times 10^6 {\rm yr}
\>, &\cr
&\approx \eta E_{SN} \nu \dot M _{\ast} t \>, \qquad t \gg 3.2
\times 10^6  {\rm yr} \>.&(5)\cr}
$$
The epoch of the onset of galactic wind, $t_{gw}$, is defined as the
time when the thermal energy of the gas exceeds the binding energy, i.e.,
$E_{th} (t_{gw}) \ge E_{bin} (t_{gw})$. [We will neglect the effects of
stellar winds here (see Gibson 1994).] We take the binding energy
of the gas as (Saito 1979; also, eqns ($1$) and ($18$) of Arimoto and
Yoshii 1987 for $g \ll 2$),
$$
E_{g, bin} \approx 1.5 \times 10^{55} ({g \over 0.1}) \Bigl ({M_T
\over 10^9 \> M_{\odot}} \Bigr )^{1.45} \quad {\rm erg} \>. \eqno(6)
$$
As noted by previous workers, the
energy accumulated in the gas in dwarf galaxies exceeds the binding 
energy of the gas even when a small fraction of the gas is turned into
stars ($M_s \la M_g$) (Larson 1974; Saito 1979; Dekel \& Silk 1986).
We have accordingly assumed here that $M_s \la M_g$. Eqns (5) and (6) 
then give,
$$
t_{gw} \approx 8.2 \times 10^{-3} n^{-1/2} g^{1/2} \Bigl 
({\eta \over 0.1} \Bigr )^{-1}  \Bigl ({\nu _{50} \over 0.35} \Bigr 
)^{-1}  ({\tau \over 0.5})^{-1} \Bigl ( { M_T \over 
10^9 \> M_{\odot} } \Bigr )^{0.45} \>{\rm Gyr}\>. \eqno(7)
$$
This time scale can also be written as 
$$
t_{gw} \approx 9 \times 10^{-3} \Bigl ({\eta \over 0.1} \Bigr )^{-1}
 \Bigl ({\nu _{50} \over 0.35})^{-1}
\Bigl ({\tau \over 0.5} \Bigr )^{-1} \Bigl ( { M_T \over 10^9 \> 
M_{\odot} }\Bigr ) ^{0.775} \> {\rm Gyr} \>, \eqno(7a)
$$
using the density of gas as indicated
by the radius from eqn ($1$). Note that $t_{gw}$ given above is the time taken
by SNRs to excite the galactic wind, {\it after} the first generation 
supernovae explode, {\it i.e.}, after $\sim 3.2 \times 10^6$ yr. 
Therefore, to compare with the values obtained numerically by previous
workers, {\it e.g.}, Yoshii \& Arimoto (1987, Fig.1), Matteucci \&
Tornamb\`e (1987, Table 1), one should add $\sim 3.2 \times 10^6$ yr to
the $t_{gw}$ from eqn (7), to find the {\it age} of the galaxy at the time
of the galactic wind. The numbers from eqn (9) and those
in the previous numerical works are consistent within a factor of a few.
However, eqn (7) should not be taken too literally for larger masses
($M_T \ga 10^{11}$ M$_{\odot}$),
for which the assumption of constant supernova rate fails. But for dwarf
galaxies ($M_T \le 10^{11}$ M$_{\odot}$) this assumption is a reasonable
one, since the time scale of the galactic wind is much
shorter for these galaxies than the time scale of the decline of the
supernovae rate after the first peak (Matteucci \& Greggio 1986).

It can also be seen from the 
supernovae rates of Type I and Type II, as computed by Matteucci \&
Greggio (1986), that SN Is do not begin to contribute substantially to
the enrichment of the gas until after a few times $10^8$ yr. This is
larger than the galactic wind time scale for galaxies of our interest.
Therefore, to calculate the metallicity of the gas expelled by the
dwarf galaxies in clusters, we will only consider Type II supernovae.

Renzini \etal (1993) compared the amount of iron produced in SNII, as
predicted by various models, and found that an average amount of $\sim 0.07$
M$_{\odot}$ of iron is produced per SN II, and that it is fairly independent
of the slope of IMF. We write the total
amount of iron in the galaxy at the time of the galactic wind as,
$M_{Fe} \sim 0.07 \nu \dot M_{\ast} t_{gw}$, using $t_{gw}$ from eqn
(9). Approximately, the stars that form {\it within} $\sim t_{gw}$
after the onset of star formation are the ones that become supernovae 
{\it before} the gas is ejected out of the galaxy, and only those stars
get to enrich the gas in the galactic wind. This yields,
$$\eqalignno{
M_{Fe} &\sim 0.07 \nu \dot M_{\ast} t_{gw} \quad (M_{\odot}) &\cr
&\sim 1.05 \times 10^4 \Bigl ({\eta \over 0.1} \Bigr )^{-1}
\Bigl ({g \over 0.1} \Bigr )
\Bigl ( { M_T \over 10^9 \> M_{\odot}} \Bigr )^{1.45}
\quad M_{\odot} \>. &(8)\cr}
$$
The average metallicity of the galaxy at the epoch of galactic wind, is
then given by [dividing eqn (8) by $gM_T$, since the total content of
iron would be shared by gas and the stars],
$$
{Z_{Fe} \over Z_{\odot,Fe}} = {M_{Fe} \over gM_T Z_{\odot,Fe}}
\sim 0.056 \left({\eta \over 0.1}\right)^{-1}
\left({M_T \over 10^9 M_{\odot}} \right)^{0.45} \>. \eqno(9)
$$
We have used $Z_{\odot, Fe}=1.89 \times 10^{-3}$ (Renzini \etal 1993).
The values predicted by the above equation match well those from Fig.1 of 
Yoshii \& Arimoto (1987) for galaxies $M_T \la 10^{11}$ M$_{\odot}$
(see also Arimoto \& Yoshii 1987, Matteucci \& Tornamb\`e 1987).
However, one should remember the
different assumptions adopted by these authors when comparing the results
(e.g., the limits on the masses of supernovae progenitors, slope of IMF etc.). 
We find that the values given by our equations are within a factor of
a few from previous numerical results. Moreover, the equations above are
more transparent with all the dependences shown explicitly, than the
numerical results.

We can estimate the gas mass in the wind following Larson (1974). 
We define the gas mass
and mass which has formed into stars, as $M_g$ and $M_s$, respectively.
The binding energy of the gas is ${E_{g,bin} \over M_g} \sim 1.5 \times
10^{47} \Bigl ({M_T \over 10^9 \> M_{\odot}} \Bigr )^{0.45}$ erg 
M$_{\odot} ^{-1}$ [see eqn (6)].
The energy in the gas can be estimated as $10^{48} ({\nu _{50} \over 0.5})
({\eta \over 0.1}) M_s$ erg, for a typical supernova energy of $10^{51}$ erg. 
This means that gas will escape from
the galaxy when ${M_g \over M_s} \sim 6.67 \Bigl ({M_T \over 10^9 
\> M_{\odot}} \Bigr )^{-0.45} ({\nu _{50} \over 0.5}) ({\eta \over 0.1})\> 
\sim \Bigl ({M_T \over 6.78 \times 10^{10} \> M_{\odot}} \Bigr )^{-0.45}
({\nu _{50} \over 0.5}) ({\eta \over 0.1})$. 
Therefore, remembering that $g M_T= M_g +M_s$, one can write the fraction of 
mass that is expelled from dwarf galaxies, as
$$
\gamma= {M_g \over M_T} \approx 0.1 \left({g \over 0.1}\right) \> { {
({M_T \over 6.78 \times 10^{10} \> M_{\odot}})^{-0.45} ({\nu _{50} \over 
0.5}) ({\eta \over 0.1}) \over 1+{({M_T \over 6.78 \times 10^{10}\>
M_{\odot}})^{-0.45} } ({\nu _{50} \over 0.5}) ({\eta \over 0.1}) }} 
\>. \eqno(10)
$$

However, this expression for $\gamma$ is incorrect at the low mass end,
because it predicts an unrealistically large value. The above expression 
neglects the fact that a lot of mass is locked up in low mass stars at the time
of the explosion of the first generation supernovae, and it underestimates the 
mass of stars at the
time of the galactic wind. To take this into account, we first note that the
galactic wind for galaxies smaller than $gM_T \la 10^7$ M$_{\odot}$ occurs
soon after the first generation of supernovae explode. We then note that the 
fraction, $m_l$, of the stellar mass ($M_s$) that is in low mass
stars, can be written in term of the IMF slope $x$, as
$$
m_l = \Bigl ( {1 \over M_l^{x-1}} - {1 \over M_{l,sn}^{x-1}}
\Bigr ) \Bigl ({1 \over M_l^{x-1}} -{1 \over M_u^{x-1}} \Bigr )^{-1}
\>. \eqno(11)
$$
For $M_l$, $M_u$, and $M_{l,sn}=0.1$, 100, and 8 M$_{\odot}$, respectively,
$m_l=0.86$ $(x=1.35)$ and $m_l=0.59$ $(x=0.95)$. Now, the gas mass 
in the galaxy decreases due to star formation as,
$$
{d \> M_g \over dt} = - \tau \left( {M_g \over t_{ff}} \right)
 = - \tau \left( {1 \over 2.1 \times 10^7 \>{\rm yr}} \right)
     \left({M_T \over 10^9 M_{\odot}}\right)^{-0.325} M_g \>, \eqno(12)
$$
where we have used the earlier expression for the star formation rate, and 
used a value of $n$ that is given by eqn (2) for ionized gas.
 The mass of gas
that remains in the galaxy after $t \sim 3.2 \times 10^6$ yr, the epoch of
the first generation supernovae ($\sim$ life time of very massive stars) is
given by,
$$
M_g' =
  gM_T \exp \left( - 0.16 \tau
     \left({M_T \over 10^9 M_{\odot}}\right)^{-0.325} \right) .
\eqno(13)
$$
However, this neglects the fact that by the time of the first generation
supernovae, the massive stars [of total mass, $(1-m_l) M_s$] would
return most of their mass back into the galaxy.
Therefore, the actual gas mass $M_g = M_g' + ( 1 - m_l) (gM_T - M_g')$.
This gives us an expression for $\gamma$ that is more realistic for
low mass galaxies,
$$
\gamma=
  g\left[1- m_l \left[ 1- \exp \left(- 0.16 \tau
     \left({M_T \over 10^9 M_{\odot}}\right)^{-0.325} \right) \right] \right].
\eqno(14)
$$
We will use a value of $\gamma$ that is smaller of the two values
predicted by eqns (10) and (14). 

The resulting $\gamma$ is plotted against galactic masses at the present time,
$M=M_T(1-\gamma)$, in Fig.1(a). It shows a maximum
around $M \sim 10^{8-9} M_{\odot}$: a large amount of gas is lost
from such typical dwarf galaxies. For larger galaxies, the deeper gravitational
potential does not allow the hot gas to escape, while for smaller galaxies,
most of the gas is converted into stars prior to galactic wind.
These values of $\gamma$ are physically more realistic than the constant
$\gamma$ for galaxies of all masses used by Trentham (1994). 

Figure 1(b) shows the corresponding iron abundance of expelled gas in units
of the solar value. Smaller galaxies with $M \sim 10^{10} M_{\odot}$ have
less time to enrich the gas ($t_{gw} \la 10^8$ yr), giving a low
metallicity, of order less than 0.5 $Z_{\odot}$. 
 If the conversion efficiency
of the blast energy into the gas, denoted as $\eta$, is lowered,
the epoch of galactic wind is prolonged and thus the metallicity is increased.
The mass of expelled iron is estimated as
$\gamma M_{T} Z_{Fe} = {\gamma \over 1-\gamma} M Z_{Fe}$, and the
resulting iron masses for $M \sim 10^{9-11} M_{\odot}$
are in good agreement with those numerically derived by Matteucci \& 
Tornamb\`e (1987) and David \etal (1990).

%%%%%%%%%%%%%%%%%%%%%%%%%%%%%%%%%%%%%%%%%%%%%%%%%%%%%%%%%%%
%%% Section 3 %%%%%%%%%%%%%%%%%%%%%%%%%%%%%%%%%%%%%%%%%%%%%%%
%%%%%%%%%%%%%%%%%%%%%%%%%%%%%%%%%%%%%%%%%%%%%%%%%%%%%%%%%%%%%
\bigskip\medskip
\centerline{3. HOT GAS AND IRON IN COMA CLUSTER}
\bigskip

Having derived the physically plausible expressions for
the ejected gas fraction $\gamma$ of initial total mass and
its iron metallicity $Z_{Fe}$ in each dSph galaxy, we estimate the total
amount of hot gas and iron in the ICM.
Following Trentham (1994), we use a Schechter luminosity
function $\phi(L)$ with the faint-end slope of $\alpha$, {\it i.e.},
$\phi (L) dL= \phi ^\ast \exp \Bigl (-{L \over L^\ast} \Bigr )
\Bigl ({L \over L^\ast} \Bigr )^{\alpha} {dL \over L^\ast}$. Also,
as in Trentham (1994), we use a
scaling law for the mass-to-light ratio, $(M/L) \propto L^{\beta}$,
for dSph galaxies. The characteristic luminosity $L^{\ast}$, at
the knee of Schechter function, is $\sim 1.42 \times 10^9 h_{50}^{-2}
L_{\odot}$ for dSph galaxies (Sandage et al. 1985).
Converting into the mass function with
$M^{\ast}/L^{\ast}=15$ for the characteristic mass (Trentham 1994)
and using $\gamma$, we derive the total mass of gas ejected from
dSph galaxies as
$$
M_{gas} = N^{\ast} M^{\ast} \int^{M_{+}}_{M_{-}}
{\gamma \over 1-\gamma}
\exp \Bigl (- \Bigl ({M \over M^\ast}
\Bigr )^{1/( \beta +1)} \Bigr ) \Bigl ( {M \over M^\ast} \Bigr )
^{( \alpha + 1 )/ (\beta +1)} {dM \over (1+ \beta ) M^\ast} \>,
\eqno(15)
$$
where $M$ denotes the present-day galaxy mass, and $N^{\ast}$ is
a characteristic number of dSph galaxies, which is equal to 302.6 for Coma
(Trentham, private communication). The integral is performed over
all masses of dSph galaxies between $M_{-}$ and $M_{+}$.
Similarly, the final metallicity of the cluster gas is
estimated as
$$
Z^{cl}_{Fe} = { \int {\gamma M Z_{Fe} \over 1- \gamma} 
\phi (M) dM \over
\int {\gamma M \over 1-\gamma}\phi (M) dM} \>,
\eqno(16)
$$
where $Z_{Fe}$ is estimated from eq.(9).

Adopting the mass of observed hot gas in Coma within 5$h_{50}^{-1}$ Mpc
as $\sim 5.1 \times 10^{14} h_{50}^{-2.5} M_{\odot}$
(Briel et al. 1992), we derived the required combinations of
($\alpha,\beta$) to reproduce all the gas in Coma by galactic winds, and 
plotted them
in Fig.2(a). The thick solid line corresponds to a low-mass bound of
$M_{-}=10^4 M_{\odot}$, while the thin line, to $M_{-}=10^6 M_{\odot}$.
We adopt $h_{50}=1$ here, but different values, such as $h_{50}=2$
give essentially the same results.
Comparing with the estimate by Trentham (1994) with the assumption of
$\gamma=const.$, our physically realistic values of $\gamma$ imply
more stringent bounds on ($\alpha,\beta$); when $M_{-}
=10^4 M_{\odot}$ and $x=0.95$, our required combinations of
($\alpha,\beta$) are similar to his minimum case of
$\gamma \sim 0.065 = const.$, and a Salpeter IMF and/or larger $M_{-}$
give smaller values for $(\alpha,\beta)$.
For instance, if the observationally allowed lower limit
for the faint-end slope of the Schechter function is $-1.9<\alpha <-1.7$
(Driver et al. 1994), Fig.2(a) implies $-0.55 < \beta < -0.37$, compared to
$-0.42 < \beta < -0.27$ in Trentham (1994). Thus it becomes more difficult to
reconcile with the observationally derived value of $\beta = -0.22
\pm 0.09$ for dSph galaxies (Kormendy 1990).

Figure 2(b) shows the resulting iron metallicity in Coma, in the
hypothesis of dwarf galaxies as sources of the intracluster gas.
The thick solid curve presents the case with parameters
($M_{-}=10^4 M_{\odot},\eta=0.1$), indicating the very low
abundance below 0.01 compared to the typically observed
value $\sim 0.2$ (Hughes et al. 1988). Changing $M_{-}$ from $10^4
M_{\odot}$ to $10^6 M_{\odot}$ (so that the gas ejection is
prefrentially from more massive galaxies)
or from $\eta=0.1$ to
0.05 (so that $t_{gw}$ is prolonged), simply doubles the iron
abundance as $0.01 \sim 0.025$ and does not produce the observed
high abundance. Thus we conclude that even if the
number of dSph galaxies can supply the whole cluster gas, they
are not the main sources of high metallicity observed in many
clusters, because of the rapid loss of gas after the
first burst of star formation. The smaller the low-mass limit
of dSph galaxies, the smaller is the iron abundance, due to the increased
contribution of smaller galaxies to ICM. The effects of
Type Ia SNe, even if they contribute, do not provide more than three
times iron of the value obtained here (Renzini et al. 1993).
Therefore, we conclude that galactic winds from normal
bright ellipticals are the main sources of intracluster
metals (e.g., Matteucci \& Vettolani 1988;
Arnaud et al. 1992; Okazaki et al. 1993).

%%%%%%%%%%%%%%%%%%%%%%%%%%%%%%%%%%%%%%%%%%%%%%%%%%%%%%%%%%%%%
%%% Section 4 %%%%%%%%%%%%%%%%%%%%%%%%%%%%%%%%%%%%%%%%%%%%%%%
%%%%%%%%%%%%%%%%%%%%%%%%%%%%%%%%%%%%%%%%%%%%%%%%%%%%%%%%%%%%%
\bigskip\medskip
\centerline{4. DISCUSSION}
%%% 4.1 %%%
\bigskip

Here we discuss some implications of our results above. First, we
discuss the origin of the intracluster gas, in the light of the 
constraints on the luminosity function of dSphs in clusters to
be able to supply the hot gas, as plotted in Fig. 2(a). We consider 
sources for the primordial gas that can explain the X-ray observations.
Then, in \S 4.2, we discuss the implications of our results on the
iron abundance of the ICM in the case of dwarf galaxies being the
dominant contributors.

\bigskip

\centerline{4.1. {\it Origin of the ICM}}
\bigskip

It has been argued by some previous workers that most of the X-ray 
emitting hot gas in clusters
ought to be primordial, because the total gas ejected from bright ellipticals
is insufficient to account for the large mass of hot gas.
Okazaki \etal (1993), for exmaple, estimated that less than 10\% of the 
observed gas mass can be explained by gas ejected from large ellipticals.
%with mass larger than total stellar mass (David et al. 1990). 
The primordial gas may have pervaded the cluster
from the beginning, supplied by later
gas infall (White et al. 1993), or gas which failed to accrete into local
galactic potentials (Evrard et al. 1994). In contrast, Trentham's (1994) idea 
is that dwarf galaxies can supply most of the cluster gas via galactic winds,
on the assumption that the fraction $\gamma$ of ejected gas mass 
is constant for all galactic masses. Here, based on more realistic estimates of
$\gamma$, we have found that in Coma, one needs smaller ($\alpha,\beta$)
than his case, {\it i.e.}, more unlikely values than allowed in previous
observations ($\alpha \ga -1.8, \beta \sim -0.22$). Although
measurements of exact values of $(\alpha,\beta)$ still elude us, this
hypothesis seems difficult to support. Moreover, as we shall explain below,
gas in galaxies with $M \la 10^6 M_{\odot}$ may not cool to form galaxies,
and this makes the situation worse (Fig. 2a). 

In the context of CDM theory, most of the baryonic gas is condensed into a 
number of small objects; as a matter of fact, the number can be very large
due to the steep slope of mass function $\sim -2$.  However, the temperature
of small-scale virialized gas is below 10$^4$ K for all $M\sim 10^6 M_{\odot}$
and $\la 1\sigma$ density contrasts of $M \sim 10^{7-8} M_{\odot}$,
for which the 
cooling rate of the gas via radiation is essentially null (in the absence of
molecular gas). Lacey \& Silk (1991) showed that protogalactic objects
with mass below $M \sim 10^{6-7} M_{\odot}$ never cool within the Hubble time,
and this lower limit of the cooled mass is not changed in the presence of 
photoionization by diffuse UV background radiation
(Chiba \& Nath 1994). Probably such very small and dark objects are observed
as Lyman $\alpha$ clouds in the intergalactic space, and the gas is stably
confined in mini CDM halos (Rees 1986), with the temperature being
kept $\sim 10^4$ K in the presence of photoionization (Efstathiou 1992).
Therefore, there may be a number of small objects whose baryonic components
have not cooled and that thus have failed to form stars.

We remark here that according to numerical simulations of cluster
formation (see the review by White 1995), small-scale structures, which
were present in earlier times, disappear due to mergings and tidal forces
in the progress of hierarchical
clustering, and the final output is a regularly structured cluster with
a smooth profile. It is thus likely that uncooled gas which was originally
condensed into small objects is dispersed during hierarchical
clustering, and heated up to the virial temperature of the cluster after
the collapse of the cluster as a whole.

If this is the case, we should add this uncooled primordial gas which is 
originally resident in small scale structures,
$M_{uncooled}$, to the cluster gas. We estimate this as, 
$$
M_{uncooled} = {N^{\ast} \over \phi^{\ast}}
 \int_{M_{lim}}^{M_{-}} {gM \over 1-g} \phi (M) dM \>,
\eqno(17)
$$
where $M_{lim}$ denotes the lower limit for the mass of virialized objects.
The dotted line in Fig.2(a) corresponds to the case of $M_{lim}=10^2 M_{\odot}$
with $M_{-}=10^6 M_{\odot}$, and is comparable to the case of
Trentham's upper limit,
$\gamma=0.33= const.$ (the maximum baryon fraction in Coma). We note that
this uncooled gas can account for about 90\% of total cluster gas in this case,
and more than 80\% even when $M_{lim}=10^4 M_{\odot}$ as plotted in Fig.3.
Therefore, in order to reconcile the idea of dSphs supplying the intracluster
gas, with the present observational lower limits
for $(\alpha,\beta)$ based on realistic estimates of $\gamma$,
it is important to consider the non-negligible
fraction of uncooled {\it primordial} gas in the mass scale
below $\sim 10^6 M_{\odot}$.
Furthermore, we note that the slope of
the corresponding galaxy mass function $\phi(M)$, given as
$(\alpha-\beta)/(\beta+1)$, is $\sim -2.2$ for $M_{lim}=10^2 M_{\odot}$,
which is roughly in agreement with CDM slope of $\sim -2$. On the other
hand, without uncooled primordial gas, the slope is $\sim -2.4$ for
$M_{-}=10^4 M_{\odot}$ and $\sim -2.6$ for $M_{-}=10^6 M_{\odot}$, which
seem to be too steep. 

Therefore, we propose the uncooled gas in small objects as an alternative 
candidate for the source of the
primordial gas that can explain most of the intracluster gas. This may also
agree with the observed trend that richer clusters hold more hot gas;
small objects below $10^6 M_{\odot}$ may be more vulnerable to destruction
in richer clusters, thereby providing more gas. It is however premature now
to test this hypothesis from numerical simulations, because the existent
calculations lack enough dynamic range (e.g., a gas particle mass in
Evrard et al. 1994 is $\sim 10^8 M_{\odot}$). More sophisticated simulations
including better resolution are required to check this possibility.

We note that Trentham (1994) used a universal value of $\alpha$ throughout the
cluster, which need not be true. The dwarf galaxies near the central
region of the cluster are more vulnerable to being torn apart by tidal
forces than the dwarfs in the outer region, and this will mean that 
$\alpha$ varies with radius in clusters. It is possible
that the dwarf galaxies contribute to the ICM gas more in the
outer regions than near the central region. Babul \& Rees (1992)
have argued that the inhibition of galactic wind from dwarf galaxies
in clusters ({\it after} the ICM has formed) may give rise to
some observed differences between dwarf galaxies in the central regions
and those in the outer regions of a cluster. 

%%% 4.2 %%%
\bigskip\medskip
\centerline{4.2. {\it Metallicity of ICM}}
\bigskip

In the previous sections, we have argued that the metallicity
of the ICM gas originates mostly in Type II supernovae, if the
dwarf galaxies are indeed the sources of the observed hot gas.
We suggest that the ratios of different elements of the ICM gas 
should provide
a test of the hypothesis of the dwarfs as main contributors of
the gas. It is interesting to note that some previous workers
(e.g., Canizares 1988; White 1991) have already inferred
the origin of the metallicity of ICM in Type II supernovae, from
the observations of the ratio of iron to oxygen. The observed
ratio of $\lbrack O/Fe \rbrack$ is known to be $(3-5)$ times
the solar value, which is indicative of Type II supernovae.
This is in contrast to the prevalent notion, from the models
of galactic winds from large ellipticals, that the bulk of the
iron is produced in Type I supernovae, where the oxygen to
iron ratio is much smaller than that in Type II supernovae. 
For example, the ratio $\lbrack O/Fe \rbrack$ predicted by Matteucci \&
Vettolani (1988) is as low as $\sim 0.9$, and in other models
(e.g., David \etal 1990) it is of the order $\sim 1.5$ (Okazaki
\etal 1990).

The dominance of Type II supernovae as the contributors to the
metallicity of the ICM gas can be increased by further interactions
between the galactic winds and the intracluster gas.
It is possible that, if the explosions that ejected the gas out
of the galaxies happen at the same epoch, then the ejected gas, now
in the ICM, may
inhibit further galactic winds from any dwarf galaxies that
may have retained some fraction of its gas to undergo 
starbursts later. The heating time scale of the intracluster
gas depends on the mechanism of heating (Sarazin 1988). If
the ejected gas is heated in a time scale shorter than that of the later
starbursts from the surviving dwarf galaxies, then, as Babul \&
Rees (1992) argued, the winds can be confined by the surrounding
hot gas. They estimated that, if $(nT)_{ICM} \ga 10^4$ cm$^{-3}$ K,
then the wind from galaxies of mass $M_T \la 3 \times 10^9$ 
M$_{\odot}$ would not extend beyond the galaxy's dark halo. The
gas in the wind in this case would cool and probably fall back
on to the core. It is possible that this effect may
inhibit the injection of metals from Type I supernovae, which
begin to enrich substantially the parent dwarf galaxies after
a few times $10^8$ yr of the first generation of stars.

Since our estimates of the 
mass of iron of the ICM from dwarf galaxies fall short for clusters
with large metallicities, it is necessary to take the large 
ellipticals into account in some cases. Although the importance of Type I 
supernovae in
enriching the gas inside and in the galactic winds from large ellipticals
has been discussed in the literature, it is still a debatable issue.
It is possible that the gas inside massive galaxies and the gas in
galactic winds are enriched by different sources.
Detections of any signature of Type II supernovae origin of the ICM
gas will not necessarily rule out contributions from large ellipticals,
if Type I supernovae are found not to be important for enrichment in
the galactc wind gas.

It is possible that in rich clusters, where tidal forces are stronger,
the dwarf galaxies do not supply most of the gas in ICM, and the metallicity
is larger because the intracluster gas is mostly enriched by large
ellipticals. Poor clusters are indeed seen to harbor smaller amount
of iron than rich clusters (e.g., Mulchaey \etal 1993).

In any case, it is almost certain that further X-ray observations of
metal lines, especially with {\it ASCA}, will be useful
in elucidating the origin of the metallicity of the ICM.

%%%%%%%%%%%%%%%%%%%%%%%%%%%%%%%%%%%%%%%%%%%%%%%%%%%%%%%%%%%%%
%%% Section 5 %%%%%%%%%%%%%%%%%%%%%%%%%%%%%%%%%%%%%%%%%%%%%%%
%%%%%%%%%%%%%%%%%%%%%%%%%%%%%%%%%%%%%%%%%%%%%%%%%%%%%%%%%%%%%
\bigskip\medskip
\centerline{5. CONCLUSION}
\bigskip

We have discussed the hypothesis that gas ejected from dwarf galaxies 
in clusters can explain the observed intracluster gas. We have derived 
analytical expressions for the fraction of ejected gas, and the metallicity
of the gas, and found good agreement with previous numerical estimates.
We have then calculated the metallicity of the resulting intracluster
gas, and found that the hypothesis of dwarf galaxies as sources of the
gas, is tenable only for clusters with very low metallicity. Additional
sources, such as galactic winds from large ellipticals, are therefore
needed to explain the existence of metals. We have not included the gas
and metal mass from large ellipticals in our calculations, but only showed
the deficiency of the hypothesis of dwarf galaxies as sources of intracluster
gas, as this work is intended as a critique of this hypothesis.

We further point out that, in CDM models, the gas in small-scale structures, 
which is dispersed and incorporated as diffuse intracluster gas by the process 
of hierarchical clustering, can account for most of the 
observed hot gas. We suggest that this process is of particular relevance
in explaining the origin of the intracluster gas, especially for rich
clusters. 

\bigskip\bigskip\bigskip
%Acknowledgments:
We are indebted to N. Arimoto, N. Trentham, and Y. Yoshii,
for valuable discussions. BBN was supported by a fellowship
from IUCAA.

%\bigskip\bigskip\bigskip
 \vfill \eject
%%%%%%%%%%%%%%%%%%%%%%%%%%%%%%%%%%%%%%%%%%%%%%%%%%%%%%%%%%%%%
%%% References %%%%%%%%%%%%%%%%%%%%%%%%%%%%%%%%%%%%%%%%%%%%%%
%%%%%%%%%%%%%%%%%%%%%%%%%%%%%%%%%%%%%%%%%%%%%%%%%%%%%%%%%%%%%
\centerline{REFERENCES}\medskip

\r{Arimoto, N., \& Yoshii, Y. 1987, A\&A, 173, 23}\par
\r{Arnaud, M., Rothenflug, R., Boulade, O., Vigroux, L., \& Vangioni-Flam,
E. 1992, A\&A, 254, 49}\par
\r{Babul, A., \& Rees, M.J. 1992, MNRAS, 255, 346}\par
\r{Briel, U.G., Henry, J.P., \& Bohringer, H. 1992, AA, 259, L31}\par
\r{Canizares, C.R., Markert, T.H., \& Donahue, M.E. 1988, in {\it
Cooling Flows in Clusters and Galaxies}, Ed. A. C. Fabian (Dordrecht:
Kluwer), p. 63}\par
\r{Chiba, M., \& Nath, B.B. 1994, ApJ, 436, 618}\par
\r{Cox, D.P. 1972, ApJ, 178, 159}\par
\r{David, L.P., Forman, W., \& Jones, C. 1990, ApJ, 359, 29}\par
\r{Dekel, A., \& Silk, J. 1986, ApJ, 303, 39}\par
\r{Driver, S., Phillipps, S, Davis, S, Morgan, I., \& Disney, M. 1994,
MNRAS, 268, 393}\par
\r{Evrard, A.E., Summers, F.J., \& Davis, M. 1994, ApJ, 422, 11} \par
%\r{Fran\c cois, P 1986, A\&A, 160, 264}\par
\r{Gibson, B.K. 1994, MNRAS, 271, L35}\par
\r{Hughes, J.P., Yamashita, K., Okumura, Y., Tsunemi, H., \& Matsuoka, M.
1988, ApJ, 327, 615}\par
\r{Kormendy, J. 1990, in {\it The Edwin Hubble Centennial Symposium:
The Evolution of the Universe of Galaxies}, Ed. Kron, R.G.
(ASP: San Francisco)} \par
\r{Lacey, C.G, \& Silk, J. 1991, ApJ, 381, 14}\par
\r{Larson, R.B. 1974, MNRAS, 166, 585}\par
\r{Matteucci, F., \& Tornamb\`e A. 1987, A\&A, 185, 51}\par
\r{Matteucci, F., \& Vettolani, G. 1988, A\&A, 202, 21}\par
\r{Mulchaey, J.S., Davis, D.S., Mushotzky, R.F., \& Burstein, D.
1993, ApJ, 401, L9}\par
\r{Okazaki T., Chiba, M., Kumai, Y., \& Fujimoto, M. 1993, PASJ, 45, 669}\par
\r{Rees, M.J. 1986, MNRAS, 218, 25P}\par
\r{Renzini, A., Ciotti, L., D'Ercole, A., \& Pellegrini, S.
1993, ApJ, 419, 52}\par
\r{Saito, M. 1979, PASJ, 31, 181}\par
\r{Sandage, A., Binggeli, B. \& Tammann, G.A. 1985, AJ, 90, 1759} \par
\r{Sarazin, C.L. 1988, {\it X-ray emissions from clusters of galaxies},
Cambridge University Press: Cambridge}\par
\r{Searle, L., \& Sargent, W.L.W. 1972, ApJ, 173, 25}\par
\r{Talbott, R.J., \& Arnett, W.D. 1971. ApJ, 170, 409}\par
\r{Tinsley, B.M. 1980, Fund. Cosmic Phys., 5, 287}\par
\r{Trentham, N. 1994, Nat, 372, 157}\par
\r{White, R.E.III 1991, ApJ, 367, 69}\par
\r{White, S.D.M., Navarro, J.F., Evrard, A.,E., \& Frnk, C.S.
1993, Nat, 366, 429} \par
\r{White, S.D.M. 1995, Lectures given at Les Houches, MPA 831
preprint}\par
\r{Yoshii, Y., \& Arimoto, N. 1987, A\&A, 188, 13}\par

\vfill \eject
%%%%%%%%%%%%%%
\noindent{\bf Figure Captions}
\medskip

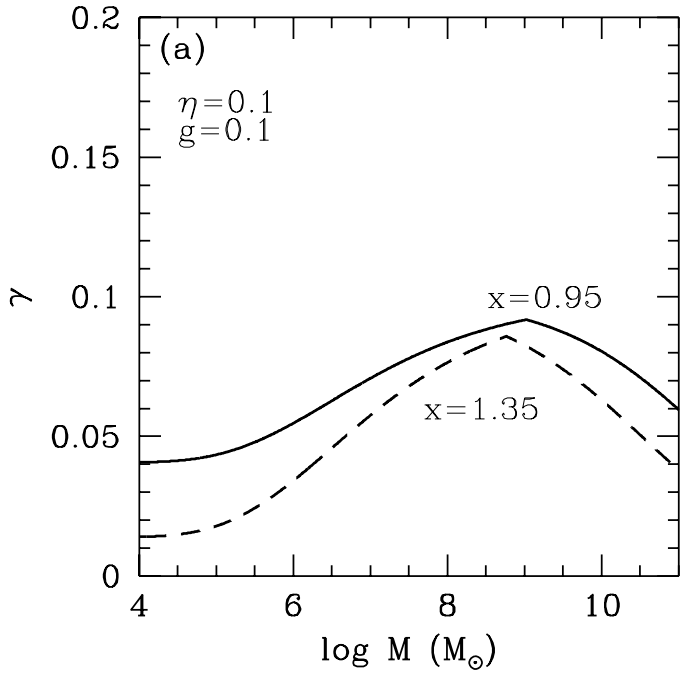
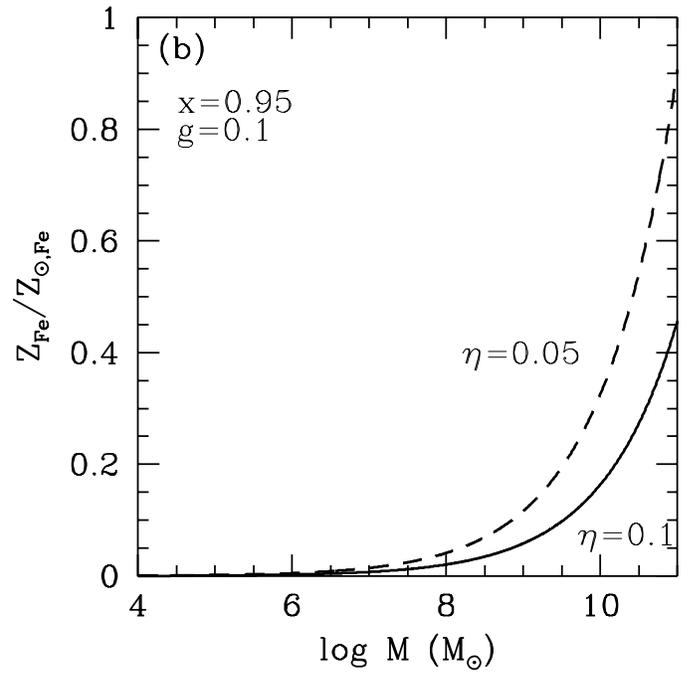
\noindent{{\bf Figure 1.} (a) Fraction of gas mass $\gamma=M_g/M_T$
expelled from galaxies as a function of the present-day galaxy mass
($\eta=0.1,\ g=0.1,\ \tau=1.0$).
{\it Thick solid curve}: for an IMF with slope index $x=0.95$.
{\it Thick dashed curve}: for $x=1.35$ (all curves are for
$M_l=0.1, M_{l,sn}=8, M_u=100 M_{\odot}$).
(b) Iron abundance of gas $Z_{Fe}/Z_{\odot,Fe}$ at the epoch of
galactic wind as a function of the present-day galaxy mass
($x=0.95,\ g=0.1,\ \tau=1.0$).
{\it Thick solid curve}: for the fraction $\eta=0.1$
of the supernova energy that is finally converted into the gas.
{\it Thick dashed curve}: for $\eta=0.05$.}

\medskip
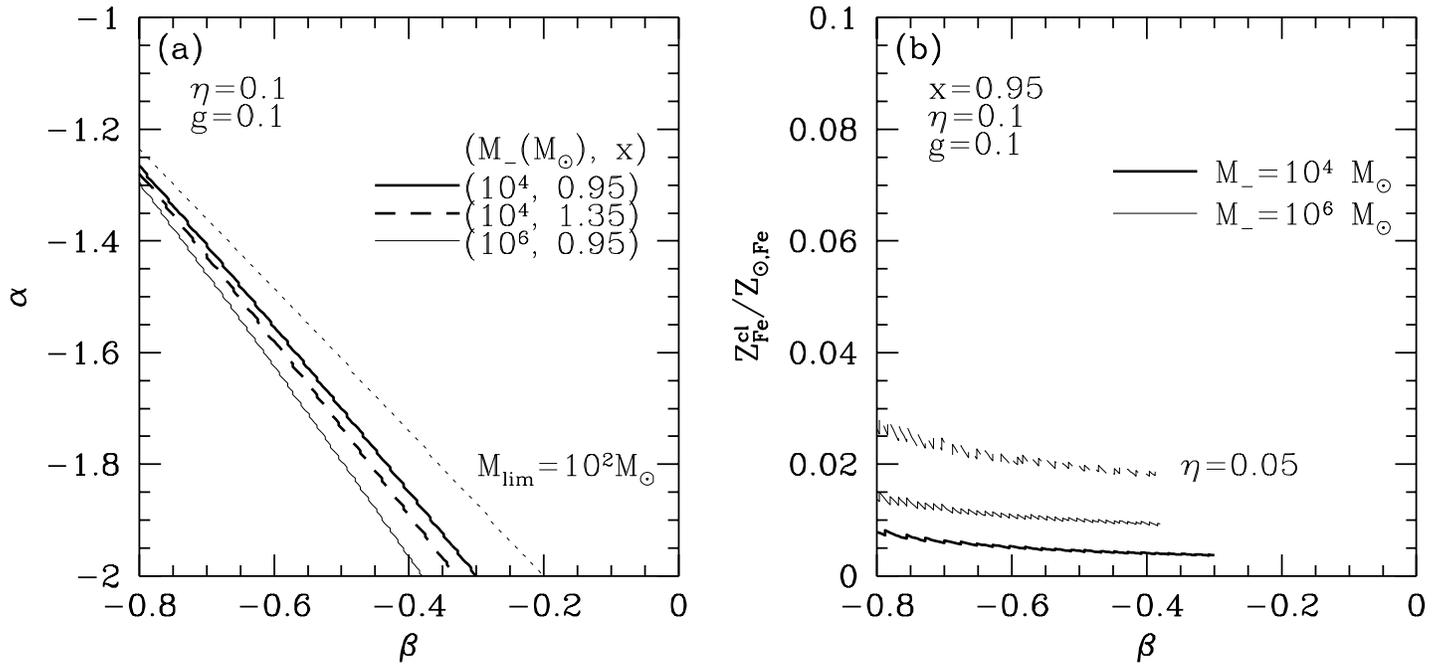
\noindent{{\bf Figure 2.} (a) Constraints on two parameters
$\alpha$ (the slope of the dSph luminosity function at faint end)
and $\beta$ [$(M/L)\propto L^{\beta}$] to reproduce the observed
hot gas in Coma by gas ejection from dSph galaxies
($h_{50}=1,\ \eta=0.1,\ g=0.1$,
the upper mass bound of dSph is $M_{+}=10^{11}
M_{\odot}$).
{\it Thick solid line}: for the lower mass bound of dSph being
$M_{-}=10^4 M_{\odot}$ and IMF index $x$ of 0.95.
{\it Thick dashed line}: for $M_{-}=10^4 M_{\odot}$ and $x=1.35$.
{\it Thin solid line}: for $M_{-}=10^6 M_{\odot}$ and $x=0.95$.
{\it Thin dotted line}: when the uncooled primordial gas in the mass range
$M_{lim}=10^2M_{\odot} \sim M_{-}=10^6M_{\odot}$ given by eq.(17)
is added to the expelled gas
plotted as {\it thin solid line}.
(b) Predicted iron abundance of ICM in Coma,
$Z^{cl}_{Fe}/Z_{\odot,Fe}$, if all of the gas mass is enriched within dSph galaxies 
($x=0.95,\ g=0.1$, $M_{+}=10^{11} M_{\odot}$).
{\it Thick solid curve}: for $M_{-}=10^4 M_{\odot}$ and $\eta=0.1$.
{\it Thin solid curve}: for $M_{-}=10^6 M_{\odot}$ and $\eta=0.1$.
{\it Thin dashed curve}: for $M_{-}=10^6 M_{\odot}$ and $\eta=0.05$.}

\medskip
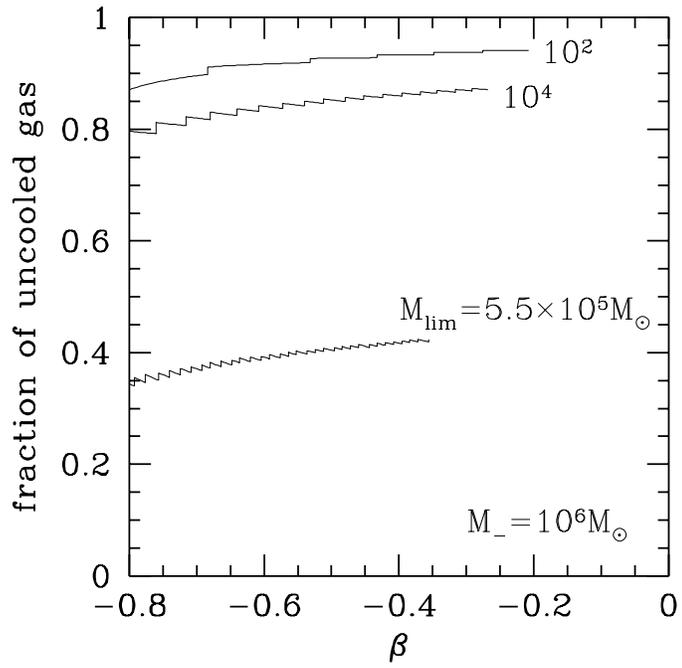
\noindent{{\bf Figure 3.} Fraction of uncooled gas
erased from low-mass objects $M_{lim} \leq M \leq
M_{-}=10^6 M_{\odot}$ in Coma ($\eta=0.1,\ g=0.1,\ x=0.95,\ M_{+}=
10^{11}M_{\odot}$).}

%%%%%%%%%%% the end %%%%%%%%%%%%%%
\vfill \eject

\end